\documentstyle[12pt,a4,epsfig]{article}
\begin{document}
\input psfig
\begin{center}
{\Large{\bf Spectroscopy of one-neutron halo nuclei from the  
Coulomb breakup reactions}}\footnote{Talk presented at the international
workshop on Production of Radioactive Ion Beams (PRORIB-2001), held
at Puri, India, Feb. 12 - 17, 2001.}\\[1.0 cm]

{\bf R. Shyam and R. Chatterjee}\\[0.5 cm]

{\it Theory Group, Saha Institute of Nuclear Physics, 1/AF Bidhan Nagar, \\
Calcutta - 700 064, INDIA.}

\end{center}

\begin{abstract}
We review the current status of obtaining the spectroscopic information on
the one-neutron halo nuclei from the Coulomb breakup reactions.
The theory of these reactions formulated in the framework of the
Distorted Wave Born Approximation, allows the use of the realistic wave
functions corresponding to any orbital angular momentum structure for
the core-valence neutron relative motion in the ground state of the
projectile. The energy, angular and parallel momentum distributions of the
projectile fragments calculated within this theory are selective
about the ground state wave function of the projectile. Therefore, 
firm conclusions can be drawn about the structure of the projectile 
ground state by comparing the calculations with the corresponding data. 
\end{abstract}

\renewcommand{\vec}[1]{\mbox{\boldmath$#1$\unboldmath}}
\newcommand{\cp}{\chi^{(+)}}
\newcommand{\cm}{\chi^{(-)*}}
\newcommand{\cmm}{\chi^{(-)}}
\newcommand{\vv}{V_{bc}({\bf r}_1)}
\newcommand{\ri}{{\bf r}_i}
\newcommand{\ro}{{\bf r}_1}
\newcommand{\ak}{{\bf k}_a}
\newcommand{\bq}{{\bf k}_b}
\newcommand{\lB}{\hat{l}\beta_{lm}}
\newcommand{\B}{\beta_{lm}}
\newcommand{\rc}{{\bf r}_c}
\newcommand{\cq}{{\bf k}_c}
\newcommand{\we}{\Psi^{(+)}_a(\xi_a,{\bf r}_1,{\bf r}_i)}
\newcommand{\bm}{\bibitem}
\newcommand{\fa}{ _2F_1(1-i\eta_a,1-i\eta_b;2;D(0))} 
\newcommand{\fB}{ _2F_1(-i\eta_a,-i\eta_b;1;D(0))}
 
\section{Introduction}

Some nuclei lying close to the neutron drip line, have one or 
two very loosely bound valence neutrons which extend too far out in
the coordinate space
with respect to the charged core~\cite{tani85}. The properties of these
neutron halo nuclei \cite{han87} have been reviewed by several authors
(see e.g.  \cite{mue93,rii94,han95,tani96}). The halo systems 
are characterized by large reaction and Coulomb
dissociation cross sections \cite{tani88,sack93,nak1,nak2}. Moreover, in
the breakup reactions induced by these nuclei, the   
angular distributions of neutrons measured in coincidence with the
core nuclei~\cite{ann90,mar96} are strongly forward peaked and  
the parallel momentum distributions of the core fragments have  
very narrow widths \cite{orr95,mex,baz,bm92,ps95,bau98}.
Due to their strikingly different properties in comparison to  
the stable systems, the halo nuclei provide a stringent test of the nuclear
structure models developed for the study of the nuclei lying close
to the line of stability. 

The Coulomb breakup, which is a significant reaction channel in the
scattering of the halo nuclei on stable heavy targets,
is a convenient tool to investigate their structure.
For instance, it would place constraints on their electric dipole response
\cite{nak1,nak2,ber88,ber91}. Of course, in the Serber \cite{serb47} 
type of models \cite{neg96,han96}, the breakup cross sections are directly
related to the momentum space wave function of the projectile ground state.
The studies of the Coulomb dissociation of the weakly bound nuclei are also of
interest due to their application in determining the cross sections of the
astrophysically interesting radiative capture reactions at solar
temperatures \cite{baur94}. The distinct advantage of the Coulomb breakup 
is that the perturbation due to the electric field of the nucleus is
known exactly.

Recently, a full quantum mechanical theory of the Coulomb breakup
reactions has been formulated \cite{cha00} within the post form distorted
wave Born approximation (DWBA). The finite range effects are included
in this theory (to be referred as FRDWBA) which can be applied to
projectiles of any ground state
angular momentum structure. In this paper, we show that definite 
information about the ground state structure of the one-neutron halo
nuclei (${\it e.g.}$, $^{11}$Be and $^{19}$C) can be obtained by 
comparing the predictions of this theory with the data on the breakup
reactions of these projectiles on heavy target nuclei. This is 
possible because the calculated energy, angular and longitudinal momentum
distributions of the projectile fragments are strongly selective about
the ground state wave function of the projectile. We also discuss
the application of this theory to the one-neutron removal reactions 
of the type A(a,b$\gamma$)X where the partial cross sections for transitions
to the excited bound states of the core are measured. These reactions
provide a more versatile tool for investigating the spectroscopy of the
halo nuclei.
 
In section 2, we present a very brief review of the post form DWBA theory of 
the Coulomb breakup reactions. The applications of the theory to
two types of the breakup reactions, A(a,b or n)X and A(a,b$\gamma$)X,  are
discussed in section 3. Our conclusions are presented in section 4.
 
\section{Formalism}
   
We consider the reaction $ a + t \rightarrow b + c + t $, where the 
projectile $a$ breaks up into fragments $b$ (charged) 
and $c$ (uncharged) in the Coulomb field of a target $t$. 
The post form DWBA $T$ - matrix for this case is
\begin{eqnarray}\label{ramp} 
 T & = & \sum_{\ell mj\mu} \langle \ell mj_c\mu_c|j\mu\rangle 
 \langle j_b\mu_bj\mu|j_a\mu_a\rangle i^\ell
 \hat{\ell}\beta_{\ell m},
\end{eqnarray}
where
\begin{equation}\label{radint}
\hat{\ell}\beta_{\ell m} =  
\int d\ro d\ri\cm_b(\bq,{\bf r})e^{-i\cq.\rc}
 \vv u_\ell(r_1){Y_{\ell m}({\hat{\bf r}}_1)}
 \cp_a(\ak,\ri),    
\end{equation}
with ${\hat \ell} = \sqrt{2\ell + 1}$. In Eq. (\ref{ramp}), $\ell$ is
the orbital angular momentum for the relative motion between $b$
and $c$. This is coupled to the spin of $c$ and the resultant $j$ is coupled
to the spin of (the inert core) $b$ to get the spin of $a$ ($j_a$).
$\chi's$ are the distorted waves for relative motions of the respective
systems ($b$ or $a$) with respect to the target. 
$\vv$ is the interaction between $b$ and $c$. The charged fragment $b$
interacts with the target by a point Coulomb interaction and hence 
$\chi^{(-)}_b({\bq},{\bf r})$ is a Coulomb distorted wave with incoming
wave boundary condition. For the pure Coulomb breakup, relative motion
wave function between the target and the uncharged fragment $c$ 
is a plane wave. The radial and angular parts of the wave function
associated with the relative motion of $b$ and $c$ in the ground state
of the projectile is given by  $u_{\ell}(r_1)$ and
${Y_{\ell m}({\hat{\bf r}}_1)}$, respectively. It should be noted that
in Eq. (\ref{radint}), the interaction of the target with the fragments
is included to all orders.

Eq. (\ref{radint}) involves a six dimensional integral which makes
its evaluation quite complicated. The problem gets further acute due
to the fact that the integrand involves three scattering waves which
have oscillatory behavior asymptotically. Therefore, approximate methods
have been used, such as the zero range approximation
(ZRA) \cite{satchler}, in which $u_\ell(r_1){Y_{\ell m}({\hat{\bf r}}_1)}$
is replaced by a delta function, or the Baur-Trautmann approximation
(BTA) \cite{BT}, where the coordinate of the projectile-target system
is replaced by that of the core-target system. Both these 
approximations lead to a factorization of the amplitude (Eq. 3) into
two independent parts, which reduces the computational complexity.
However, both these methods have limitations and their application to
the reactions of the halo nuclei is questionable.
     
In the FRDWBA theory, the Coulomb distorted wave of particle $b$ in the 
final channel is written as 
\begin{eqnarray} \label{lma}
\chi^{(-)}_b(\bq,{\bf r}) & = & e^{-i\alpha{\bf K}.\ro}
                           \chi^{(-)}_b(\bq,\ri). 
\end{eqnarray}
Eq. (\ref{lma}) represents an exact Taylor series expansion about
${\bf r}_i$ if 
${ {\bf K}}( = -i\nabla_{{\bf r}_i})$ is treated exactly. However,
instead of doing this we employ a local momentum approximation, where the 
magnitude of the local momentum ${\bf K}$ is taken to be 
\begin{eqnarray} \label{lmv} 
{ {K}}(R) = {\sqrt {{2m\over \hbar^2}(E - V(R))}}.
\end{eqnarray}
In Eq. (\ref{lmv}), $m$ is the reduced mass of the $b-t$ system,
$E$ is the energy of particle $b$ relative to the target in the
c.m. system and $V(R)$ is the  Coulomb potential
between $b$ and the target at a distance $R$. Therefore, 
the local momentum ${{\bf K}}$ is evaluated at some distance
$R$, and its magnitude is held fixed for all the values of ${\bf r}$. 
As is discussed in \cite{cha00}, the results of our calculations
are almost independent of the choice of the direction the local momentum.
Therefore, we have taken the directions of ${{\bf K}}$ and
${\bf k_b}$ to be the same in all the calculations presented in this
paper. For more details, we refer to \cite{cha00}.  

Substituting Eq. (\ref{lma}) into Eqs. (\ref{ramp}, \ref{radint}) we get the
following factorized form of the reduced amplitude  
\begin{eqnarray}\label{famp}
{\hat \ell}\beta^{FRDWBA} _{\ell m}& = &
\langle e^{i(\gamma\cq - \alpha {\bf K}).\ro}|V_{bc}|
\phi^{\ell m}_a(\ro)\rangle 
        \langle \chi^{(-)}_b(\bq,\ri) e^{i\delta\cq.\ri}|
\cp_a(\ak,\ri)\rangle.
\end{eqnarray}

Recently, a theory of the Coulomb breakup has been developed within an 
adiabatic (AD) model \cite{toste98,jal97}, where one assumes that the 
excitation of the projectile is such that the relative energy ($E_{bc}$)
of the $b-c$ system is quite small as compared to the total incident energy,
which allows $E_{bc}$ to be replaced by the constant separation energy
of the fragments in the projectile ground state. It was shown in \cite{jal97}
that under these conditions the  wave function $\we$
has an exact solution which when substituted in the post form T-matrix
leads to its factorization in the same way as Eq. (\ref{famp}).
This amplitude differs from that of the FRDWBA only in
the form factor part (first term), which is evaluated here at the momentum 
transfer $(\cq - \alpha \ak)$. It has been shown \cite{cha00,shy01} that
for the breakup of projectiles with very small one-neutron separation
energies and a $s$-wave neutron-core relative relative motion in its ground
state, the results of the adiabatic model and the FRDWBA are almost
identical to each other. Although the adiabatic model does not make the weak
coupling approximation of the DWBA, yet it necessarily requires one
of the fragments (in this case $c$) to be chargeless. In contrast, 
the FRDWBA can, in principle, be applied to cases where both
the fragments $b$ and $c$ are charged \cite{shyam85}. 

\section{Applications to the breakup reactions of $^{11}$Be
and $^{19}$C} 

\subsection{Structure models of $^{11}$Be and $^{19}$C}

For the ground state of $^{11}$Be, we have considered the following 
configurations :

(a) a $s$ -- wave valence neutron coupled to the $0^+$  $^{10}$Be core
[$1s_{1/2}\nu \otimes ^{10}$Be$(0^+)$] with a one-neutron separation
energy ($S_{n-^{10}Be}$) of 504 keV and a spectroscopic
factor (SF) of 0.74 \cite{jon},

(b) a $d$ -- wave valence neutron coupled to the $2^+$ $^{10}$Be core
[$0d_{5/2}\nu \otimes ^{10}$Be$(2^+)$] with a one-neutron separation
energy of 3.872 MeV (which is the sum of energy of the excited 2$^+$ core
(3.368 MeV) and $S_{n-^{10}Be}$) with a SF of 0.17, and

(c) an admixture of these two configurations with the SF values 
of 0.74 and 0.20 respectively \cite{aum00}.

In each case, the single particle wave function is constructed by
assuming the valence neutron-$^{10}$Be interaction to be of the
Woods-Saxon type whose depth is adjusted to reproduce the corresponding
value of the binding energy with fixed values of the radius
and the diffuseness parameters (taken to be  1.15 fm and 0.5 fm,
respectively). For configuration (a), this gives a potential depth
of 71.03 MeV, a root mean square (rms) radius for the valence neutron
of 6.7 fm, and a rms radius for $^{11}$Be of 2.91 fm when the size of
the $^{10}$Be core is taken to be 2.28 fm \cite{jal96}. 

For the case of $^{19}$C, there is a large uncertainty in the value of
the last neutron separation energy ($S_{n-^{18}C}$) with quoted values
varying between 0.160 MeV -- 1.0 MeV \cite{nak2,val01}. 
Most of the available data on the Coulomb dissociation of $^{19}$C
can be satisfactorily explained within the finite range DWBA theory
with the one-neutron separation energy of 530 keV \cite{cha00}.
The relativistic mean field (RMF) \cite{rmf} as well
as shell model (with Warburton-Brown effective interaction) \cite{war}
calculations predict the spin-parity of the ground state of this nucleus
to be $1/2^{+}$.

We have examined two possibilities for the ground state structure of
$^{19}$C:

(i) different binding energies (530 keV and 160 keV) with the same
configuration for the $^{19}$C ground state
[$1s_{1/2}\nu \otimes ^{18}$C$(0^+)$], and

(ii) different configurations
[$1s_{1/2}\nu \otimes ^{18}$C$(0^+)$] and
[$0d_{5/2}\nu \otimes ^{18}$C$(0^+)$]
with the same binding energy (530 keV) for the $^{19}$C ground state.

A spectroscopic factor of 1.0 has been used for each of the configuration,
and the corresponding single particle wave functions have been 
constructed in the same way as in the case of $^{11}$Be.
 
\subsection{A(a,b or n)X type of reactions}

\begin{figure}[ht]
\begin{center}
\mbox{\epsfig{file=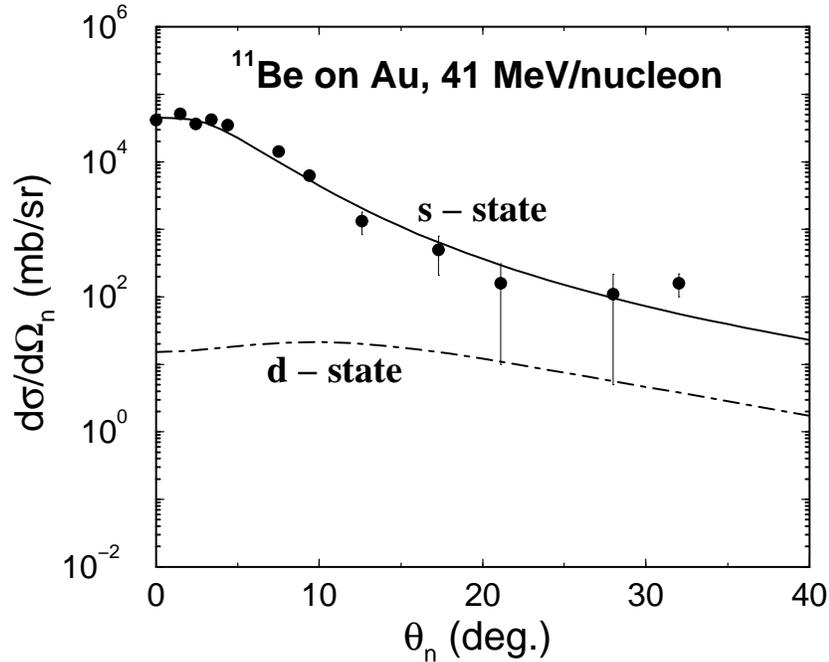,width=.75\textwidth}}
\end{center}
\vskip -0.3in
\caption{\small{The calculated neutron angular distributions for 
the breakup of $^{11}$Be on a Au target at the beam energy of 41 MeV/nucleon.
The solid and the dot-dashed lines show the results of the FRDWBA calculation
performed with the pure $s$-wave and $d$-wave configurations for the  
ground state of $^{11}$Be, respectively.} 
} 
\label{fig:figa}
\end{figure}
The measured neutron angular distribution in the exclusive
$^{11}$Be + $A$ $\rightarrow$ $^{10}$Be + n +$A$ reaction below the
grazing angle is shown to be \cite{cha00} dominated by the Coulomb
breakup process. This reflects the narrow width of the transverse momentum
distribution of the valence neutron. The neutron halo structure is also
reflected in the narrow widths of the parallel momentum distribution
(PMD) of the charged breakup fragments emitted in breakup reactions
induced by the halo nuclei.
This is because the PMD has been found to be least affected
by the reaction mechanism \cite{orr95,mex,baz,bm92,ps95} and therefore, 
a narrow PMD can be related to a long tail in the  matter distribution
in the coordinate space via Heisenberg's uncertainty principle.

In Fig. 1, we compare
the calculated and measured exclusive neutron angular distribution
$d\sigma/d\Omega_n$ as a function of the neutron angle $\theta_n$ for the 
above reaction on a Au target at the beam energy of 41 MeV/nucleon. 
Calculations are shown for both configurations (a) (solid line) and 
(b) (dot-dashed line) for the $^{11}$Be ground state. It is clear that
experimental angular distributions are reproduced by configuration (a)
only. 
\begin{figure}[ht]
\begin{center}
\mbox{\epsfig{file=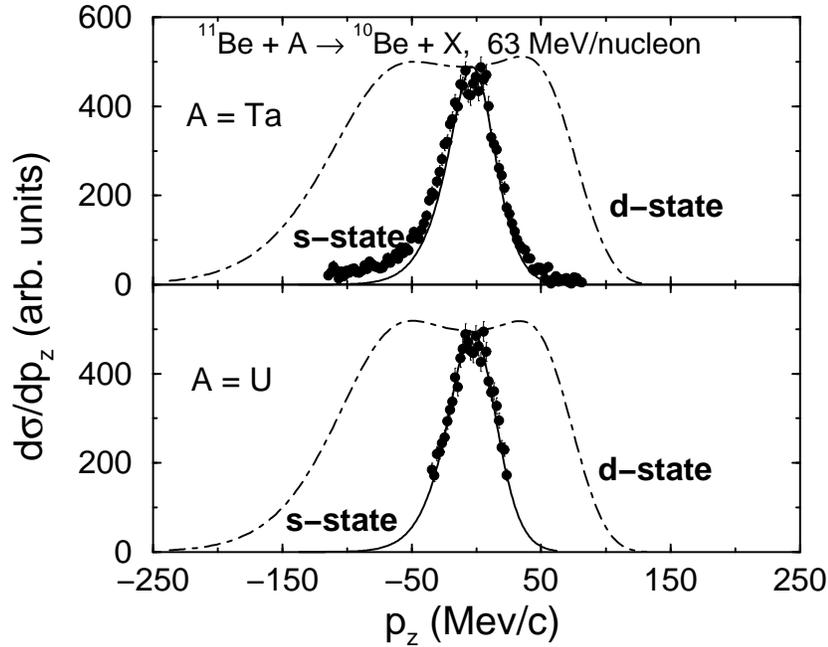,width=.75\textwidth}}
\end{center}
\vskip -0.3in
\caption{\small{
Parallel momentum distributions of $^{10}$Be in the breakup of 
$^{11}$Be on U (top half) and Ta (bottom half) at the beam energy of 
63 MeV/nucleon,  
in the rest frame of the projectile, with $s$-wave (solid line) and
$d$-wave (dot-dashed line) configurations for the $^{11}$Be ground state.
The same normalization has been used for both the cases. The data are
taken from \protect\cite{mex}.}
} 
\label{fig:figb}
\end{figure}
\noindent
In Fig. 2, we present the PMD of the $^{10}$Be fragment emitted in the
breakup of $^{11}$Be on U and Ta targets at the beam energy of
63 MeV/nucleon.  
Calculations performed with both the configurations (a) and (b) 
are shown in this figure.  The calculated cross sections are normalized
to the peak values of the data points in both the cases. While the full widths
at half maximum (FWHM) of the distributions calculated with the $s$-wave
configuration are narrow for both the targets (44 MeV/c and 43 MeV/c for
the U and Ta, respectively), they turn out to be quite broad with the
$d$-wave configuration. The narrow widths agree well with the averaged 
experimental value of 43.6$\pm$1.1 MeV/c \cite{mex}. Therefore, the
configuration (a) is favored by this data too.
The very narrow widths of the parallel momentum
distributions signal the presence of a neutron halo structure in 
$^{11}$Be.
\begin{figure}[ht]
\begin{center}
\mbox{\epsfig{file=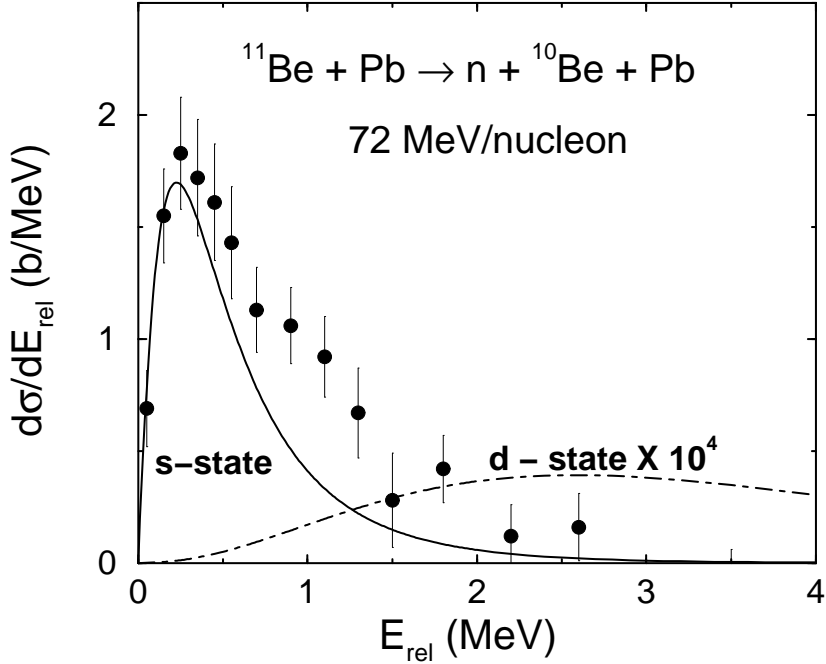,width=.75\textwidth}}
\end{center}
\vskip -0.3in
\caption{\small{
Relative energy spectra for the Coulomb breakup of $^{11}$Be on a
Pb target at the beam energy of 72 MeV/nucleon 
with the $s$-wave (solid line) and
the $d$-wave (dot-dashed line) configurations for the $^{11}$Be ground state. 
The data are taken from \protect\cite{nak1}.}} 
\label{fig:figc}
\end{figure}

The relative energy spectrum for the breakup of $^{11}$Be on a Pb target
at the beam energy of 72 MeV/nucleon is shown in Fig. 3.
The results obtained with both the
configurations [(a) and (b)] for the ground state of $^{11}$Be
are shown.  We see that, while the calculations done with the $s$-wave
configuration reproduce the data well near the peak region, the 
$d$-wave configuration underpredicts the data by several orders of 
magnitude.  However, the data at higher relative energies can not be
explained by our calculations. This can be attributed to the fact that
the nuclear breakup effects, which can contribute substantially
\cite{dasso99,typ} at the higher relative energies 
(for $E_{rel}$ $>$ 0.6 MeV),
are not included in these calculations. Of course, the authors of Ref.
\cite{nak1} claim
that their data have been corrected for these  contributions. However,
the procedure adopted by them for this purpose may not be adequate
\cite{cha00}. For a proper description of the data, 
both Coulomb and nuclear breakup contributions should be calculated
on the same footing and corresponding amplitudes should be added coherently
to get the cross sections. 

Therefore, for $^{11}$Be, a $s$ -- wave configuration
[$1s_{1/2}\nu \otimes ^{10}$Be$(0^+)$] spectroscopic factor of 0.74
for its ground state 
provides best agreement with the experimental data in all the cases. Our
calculations cannot distinguish between this and configuration (c) of
section 3.1.
\begin{figure}[ht]
\begin{center}
\mbox{\epsfig{file=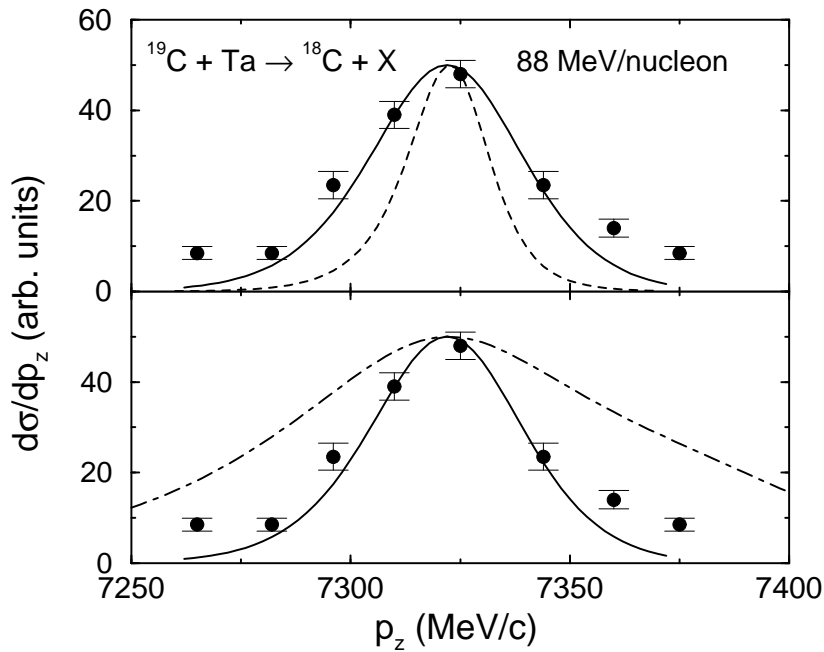,width=.75\textwidth}}
\end{center}
\vskip -0.3in
\caption{\small{FRDWBA results for the parallel momentum distribution
of $^{18}$C in the breakup of $^{19}$C on Ta target at the beam energy
of 88 MeV/nucleon. The top half shows the results obtained with the
configuration [$1s_{1/2}\nu \otimes ^{18}$C $(0^+)$] 
and single particle
wave function for the ground state of $^{19}$C with one-neutron
separation energies of 530 keV (solid line), 160 keV (dashed line).
The bottom half shows the result obtained with the configurations
[$1s_{1/2}\nu \otimes ^{18}$C $(0^+)$] (solid) and
[$0d_{5/2}\nu \otimes ^{18}$C $(0^+)$] (dot-dashed), 
with the same value of the one-neutron separation energy (530 keV).
The data have been taken from \protect\cite{baz}.}} 
\label{fig:figd}
\end{figure}

In Fig. 4, we present the PMD (calculated within the FRDWBA formalism) 
of the $^{18}$C fragment in the breakup of $^{19}$C on a Ta target at the
beam energy of 88 MeV/nucleon. We have normalized
the peaks of the calculated PMDs to that of the data (given in arbitrary units)
\cite{baz} (this also involves coinciding the position of maxima of the
calculated and experimental PMDs). As can be seen from the upper part of
this figure, the experimental data clearly favor $S_{n-^{18}C}$ = 0.53 MeV
with the $s$ -- wave n-$^{18}$C relative motion in the ground state of
$^{19}$C. 

In the lower part of Fig. 4, we have shown the results obtained
with the $d$ -- wave relative motion for this system
(with $S_{n-^{18}C}$ = 0.53 MeV) and have compared it with that
obtained with a $s$ -- wave relative motion
with the same value of the binding energy. As can be seen,
the FWHM of the experimental PMD is grossly over-estimated by the $d$ -- wave
configuration.  The calculated FWHM with the $s$ -- state configuration
(with $S_{n-^{18}C}$ = 530 keV) is 40 MeV/c, which is in excellent agreement
with the experimental value of 41$\pm$3 MeV/c \cite{baz}. Thus these data
favor a configuration
[$1s_{1/2}\nu \otimes ^{18}$C$(0^+)$],
with a one-neutron
separation energy of 0.530 MeV for the ground state of $^{19}$C. 
The narrow width of the PMD provides support to the presence of a 
one-neutron halo structure in $^{19}$C. 
\begin{figure}[ht]
\begin{center}
\mbox{\epsfig{file=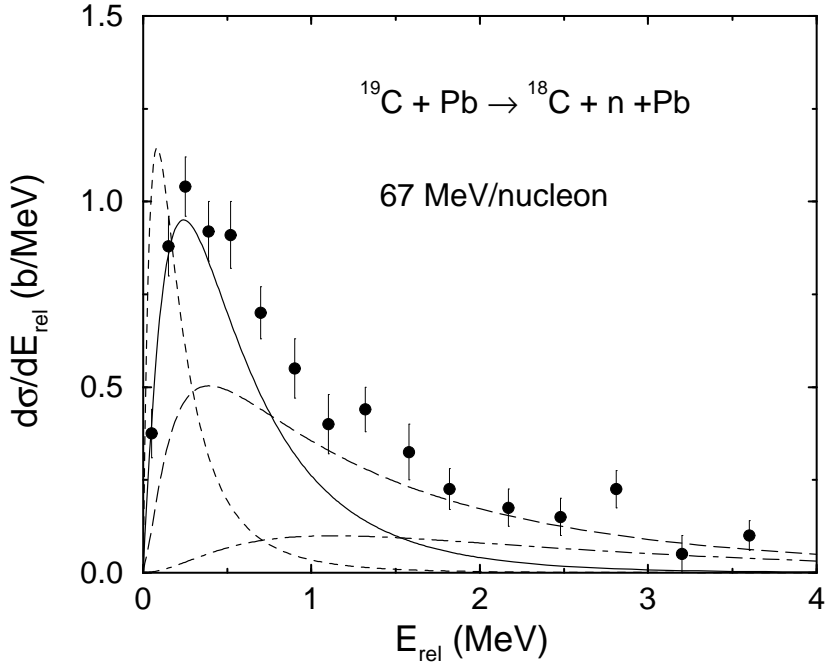,width=.75\textwidth}}
\end{center}
\vskip -0.3in
\caption{\small{
Calculated relative energy spectra for the Coulomb breakup of $^{19}$C
on Pb at 67 MeV/nucleon. The $d$ -- state results, for binding energies 160 keV 
(long dashed) and 530 keV (dot-dashed), are multiplied by 10. The 160 keV 
$s$ -- state result (short dashed) is multiplied by 0.086. The 530 keV $s$ -- 
state result is represented by the solid line. The data have been
taken from \protect\cite{nak2}.}} 
\label{fig:fige}
\end{figure}

In Fig. 5, we have shown the results of our FRDWBA calculations for the  
the relative energy spectrum for the 
breakup of $^{19}$C on Pb at the beam energy of 67 MeV/nucleon.
The experimental
data is taken from \cite{nak2}. The angular integration for the $^{19}$C
center of mass is done up to the grazing angle of 2.5$^\circ$. It can be 
seen that in this case also the best agreement with the data (near the
peak position) is obtained with the $s$ -- wave configuration with
$S_{n-^{18}C}$ = 0.53 MeV. Calculations done 
with the $d$ -- state configuration for both 530 
keV and 160 keV one-neutron separation energy fails to reproduce
the data. At the same time, those performed with the $s$ -- state
configuration but with $S_{n-^{18}C}$ = 0.16 MeV overestimates the data 
by at least an order of magnitude and also fail to reproduce its 
peak position. However, as in the case of $^{11}$Be, the calculations
underestimate the relative energy spectrum for larger values of relative
energies, and the proper consideration of the nuclear breakup effects is
necessary to explain the data in this region. 

Therefore, the results for the PMD of $^{18}$C 
and the relative energy spectrum of the $n$ + $^{18}$C system
are described best with a ground state configuration of $^{19}$C of 
[$1s_{1/2}\nu \otimes ^{18}$C$(0^+)$],
with a one-neutron separation
energy of 530 keV and a spectroscopic factor of 1.

\subsection{A(a,b$\gamma$)X type of reaction}

In the (a,b$\gamma$) type of reactions, one nucleon (usually the
valence or halo) is removed from the projectile (a) in its breakup
reaction in the field of a target nucleus and the states of the
core fragment (b) populated in this reaction are identified by their
gamma ($\gamma$) decay. The intensities of the decay photons are used
to determine the partial breakup cross sections to different core states.
The signatures of the orbital angular momentum $\ell$ associated with
the relative motion of core states with respect to the valence nucleon
(removed from the projectile) are provided by the measured parallel
momentum distributions of the core states (see Fig. 2, 4). This provides
a new and more versatile technique for investigating the spectroscopy of
nuclei near the drip line \cite{aum00,val01,nav98}.
It mixes the power of the conventional transfer reaction (sensitivity to
the $\ell$ value of the removed neutron) with the practical advantage of
the breakup reactions (larger cross sections). 
\begin{figure}[ht]
\begin{center}
\mbox{\epsfig{file=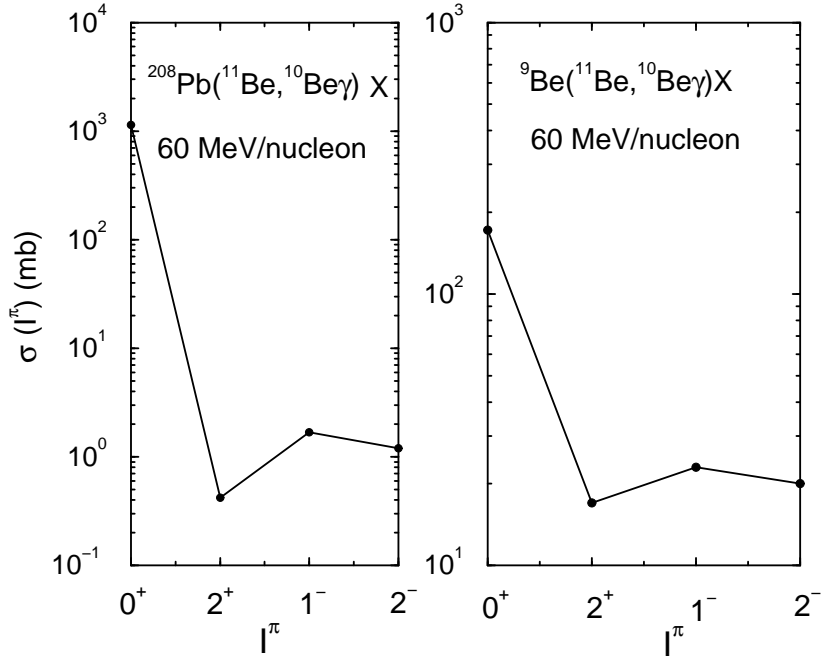,width=.75\textwidth}}
\end{center}
\vskip -0.3in
\caption{\small{
Partial total cross sections for the ($^{11}$Be,$^{10}$Be$\gamma$)
reaction on $^{208}$Pb (left) and $^{9}$Be (right) targets
at the beam energies of 60 MeV/nucleon. The states of the $^{10}$Be
core populated in each case are $0^+$, $2^+$, $1^-$ and $2^-$. The results
for the $^9$Be target are taken from ~\protect\cite{aum00}.
}}
\label{fig:figf}
\end{figure}

Most of the studies of the (a,b$\gamma$) reaction
performed so far involve a light $^9$Be target, where the breakup
process is governed almost entirely by only the nuclear interaction
between the projectile fragments and the target. It would be
interesting to know if there are quantitative differences in the
${\it relative}$ populations of the core states in the pure Coulomb
breakup mechanism, as compared to those observed in the pure nuclear
breakup process. In Fig. 6, we compare the pure Coulomb partial cross
sections for the $^{11}$Be breakup on a $^{208}$Pb target (shown in the
left panel) with the pure nuclear ones (shown in the right panel)
corresponding to the breakup of $^{11}$Be on a  $^9$Be target (taken
from \cite{aum00}) at the beam energy of 60 MeV/nucleon. Both the
cross sections are calculated for transitions to the  four $^{10}$Be
final states.  The ground (0$^+$) and excited (3.368 MeV) (2$^+$)
states were assumed to correspond to the configurations
[1$s_{1/2}\nu$ $\otimes$ $^{10}$Be($0^+$)] and
[0$d_{5/2}\nu$ $\otimes$ $^{10}$Be($2^+$)], respectively. 
The corresponding SF values for these two configurations were
taken to be 0.74 and 0.20, respectively, The excited 1$^-$ (5.956 MeV)
and 2$^-$ (6.256 MeV) states were assumed to stem from the configurations
[0$p_{3/2}\nu$ $\otimes$ $^{10}$Be($1^-$)] and
[0$p_{3/2}\nu$ $\otimes$ $^{10}$Be($2^-$)], respectively,
with the SF values of 0.69 and 0.58, respectively.
In each case, the neutron single particle wave function is calculated in
the same way as described in section 3.1.

It is evident that
in the case of pure Coulomb breakup of a projectile with a halo
ground state, most of the cross section goes to the ground state
(0$^+$) of the core. The sum of the partial cross sections corresponding
to all the excited states is less than 1$\%$ of that to the ground state.
This is in sharp contrast to the case of the lighter target where partial
cross sections corresponding to all the excited states represent about
22$\%$ of the total. Similar results have been obtained for the $^{19}$C
projectile (see \cite{shy01} for more details).
Indeed, in the measurements \cite{gui00} of the (a,b$\gamma$) type of
reactions with $^{14}$B projectile on $^{197}$Au gold target at the beam
energy of 60 MeV/nucleon no core-excited transitions were seen. Therefore,
this lends support to our observation. The transitions to the excited states
of the core corresponding to the non-zero $\ell$-values of the neutron-core
relative motion, are quite weak in the pure Coulomb breakup reaction as
compared to those in the nuclear breakup process. Therefore, (a,b$\gamma$)
type of reaction on a heavy target is potentially a more useful tool for
investigating the properties of the ground state of the core fragments.

The suppression of the cross sections to the higher states can be
understood from the strong dependence of the Coulomb breakup cross
sections on the one-neutron separation energy (SE).
In the case of the nuclear breakup, the dependence of
the cross section on SE is comparatively weaker. This could be
understood from the fact that nuclear breakup cross sections are
sensitive to the $b-c$ relative wave functions at shorter distances
which do not change much with changes in the value of SE.

\section{Concluding remarks}

Finite range DWBA theory of the Coulomb breakup reactions provides
a very powerful tool to probe the ground state structure of the
one-neutron halo nuclei. Its predictions for most of the  
breakup observable are quite sensitive to the ground state configuration
of the projectile. Therefore, their comparison with the experimental
breakup data can be used to put constraints on the ground state structure
of the one-neutron halo nuclei.

The Coulomb dominated one-neutron removal reaction of the type
(a,b$\gamma$) populates predominantly the ground state of the 
core fragment. Therefore, such studies would be very useful in
investigating the ground state structure of the halo nuclei for
those partitions in which the core fragment remains in its ground state. 

Useful discussions
with Prabir Banerjee, Pawel Danielewicz, Gregers Hansen, and Stefan Typel
are gratefully acknowledged.

\end{document}